\documentclass[12pt,preprint]{aastex}
\usepackage{times,color}

\usepackage{amsmath}

\newcommand{\bnabla}{\boldmath\mbox{$\nabla$}\unboldmath}
\newcommand{\bvel}{\boldmath\mbox{$v$}\unboldmath}
\newcommand{\bmag}{\boldmath\mbox{$B$}\unboldmath}

\shorttitle{TWO-FLUID MHD SIMULATIONS OF CONVERGING HI FLOWS}
\shortauthors{T. INOUE \& S. INUTSUKA}

\begin{document}

\title{
TWO-FLUID MHD SIMULATIONS OF CONVERGING HI FLOWS IN THE INTERSTELLAR MEDIUM. I: METHODOLOGY AND BASIC RESULTS
}
\author{Tsuyoshi Inoue\altaffilmark{1,2}, and Shu-ichiro Inutsuka\altaffilmark{1,3}}
\altaffiltext{1}{Department of Physics, Graduate School of Science, Kyoto University, Sakyo-ku, Kyoto 606-8588, Japan; tsuyoshi@tap.scphys.kyoto-u.ac.jp}
\altaffiltext{2}{Division of Theoretical Astronomy, National Astronomical Observatory of Japan, Osawa, Mitaka 181-8588 Japan}
\altaffiltext{3}{Kavli Institute for Theoretical Physics, University of California, Santa
Barbara, CA 93106}

\begin{abstract}
We develop an unconditionally stable numerical method for solving the coupling between two fluids (frictional forces/heatings, ionization, and recombination), and investigate the dynamical condensation process of thermally unstable gas that is provided by the shock waves in a weakly ionized and magnetized interstellar medium by using two-dimensional two-fluid magnetohydrodynamical simulations.
If we neglect the effect of magnetic field, it is known that condensation driven by thermal instability can generate high density clouds whose physical condition corresponds to molecular clouds (precursor of molecular clouds).
In this paper, we study the effect of magnetic field on the evolution of supersonic converging HI flows and focus on the case in which the orientation of magnetic field to converging flows is orthogonal.
We show that the magnetic pressure gradient parallel to the flows prevents the formation of high density and high column density clouds, but instead generates fragmented, filamentary HI clouds.
With this restricted geometry, magnetic field drastically diminishes the opportunity of fast molecular cloud formation directly from the warm neutral medium, in contrast to the case without magnetic field.
\end{abstract}


\section{Introduction}
It is widely known that radiative cooling and heating in the interstellar medium (ISM) leads to a stable cold dense medium and a warm diffuse medium (Field, Goldsmith \& Harbing 1969; Wolfire et al. 1995, 2003).
These stable phases are called the cold neutral medium (CNM), and the warm neutral medium (WNM).
The CNM is observed as HI clouds ($n \sim 10^{1-2}$ cm$^{-3}$, $T\sim 10^{2}$ K), and cold neutral clouds that contain substantial amounts of H$_{2}$ are called molecular clouds ($n \sim 10^{2-3}$ cm$^{-3}$, $T\sim 10$ K).
Since molecular clouds are the sites of star formation, the knowledge of how molecular clouds are formed is crucial to understanding the initial conditions of star formation.

Thermal instability (TI) is the most promising formation mechanism of the CNM.
Many authors studied the dynamical condensation process of the ISM driven by TI.
Koyama \& Inutsuka (2000, 2002), and Inutsuka, Koyama, \& Inoue (2005) studied the CNM formation as a result of TI in the layer compressed by shock propagation.
Hennebelle \& P\'erault (1999), Audit \& Hennebelle (2005), Hennebelle \& Audit (2007), Heitsch et al. (2005, 2006), and Vazquez-Semadeni et al. (2006, 2007) studied an analogous process in supersonic converging flows, which essentially include two shocks.
These studies showed that TI can generate molecular clouds, by piling up the WNM.

However, the above-mentioned studies do not include the effect of magnetic field.
The dynamical formation process of the CNM with the effect of magnetic field is studied by Hennebelle \& P\'erault (2000) using one-dimensional, ideal MHD simulations.
They showed that magnetic pressure prevents TI, if the initial angle between the magnetic field and the fluid velocity is larger than $20^\circ$-$30^\circ$ with a few microgauss initial field strength.
Recently, Inoue, Inutsuka, \& Koyama (2007) have shown that if the effect of ambipolar diffusion is taken into account, TI can create the CNM even with a microgauss magnetic field which is orthogonal to the orientation of the condensation.

In Inoue, Inutsuka, \& Koyama (2007), we used a thermally unstable equilibrium as an initial condition and performed one-dimensional numerical simulations.
However, thermally unstable gas is likely to be supplied by the shock compression of the ISM, and in multiple dimensions, growth of TI along the magnetic field seems to be faster than the mode that is perpendicular to the magnetic field.
In this paper, to investigate how the CNM is formed under more realistic situations, we study the dynamical condensation process of thermally unstable gas that is supplied by the shock compression of the WNM employing two-dimensional two-fluid magnetohydrodynamical simulations.

This paper is organized as follows.
In \S 2 we provide the basic equations of the weakly ionized interstellar medium and assumptions used in this paper.
In \S 3 we describe a method for solving the basic equations numerically, in which an unconditionally stable scheme that solves the coupling between two fluids is proposed and tested.
The results of two-fluid simulations of TIs in weakly ionized media are shown in \S 4.
Finally, in \S 5 we summarize our results and discuss their implications. 

\section{The Model}
\subsection{Basic Equations}
We consider the equations of two-dimensional hydrodynamics (HD) and magnetohydrodynamics (MHD) that include the effects of ionization, recombination, and collisional coupling between neutral and ionized gases (Draine 1986).
The effects of radiative cooling, heating, and thermal conduction are taken into account in the neutral fluid part.
\begin{eqnarray}\label{E1}
&& \frac{\partial \rho_{\rm n}}{\partial t}+\bnabla\cdot(\rho_{\rm n}\,\bvel_{\rm n})=
m_{\rm n}\,(\alpha\,n_{\rm p}^{2}-\xi\,n_{\rm n}),\\
&& \frac{\partial \rho_{\rm p}}{\partial t}+\bnabla\cdot(\rho_{\rm p}\,\bvel_{\rm i})=
m_{\rm p}\,(\xi\,n_{\rm n}-\alpha\,n_{\rm p}^{2}),\\
&& \frac{\partial \rho_{\rm n}\bvel_{\rm n}}{\partial t}+\bnabla\cdot(p_{\rm n}+\rho_{\rm n}\,\bvel_{\rm n}\otimes \bvel_{\rm n})=(A_{\rm p}\,\rho_{\rm p}+A_{\rm CI\hspace{-.1em}I}\,\rho_{\rm CI\hspace{-.1em}I})\,\rho_{\rm n}\,(\bvel_{\rm i}-\bvel_{\rm n}),\\
&& \frac{\partial\rho_{\rm i}\bvel_{\rm i}}{\partial t}+\bnabla\cdot\left(p_{\rm n}+\frac{B^{2}}{8\pi}+\rho_{\rm i}\,\bvel_{\rm i}\otimes \bvel_{\rm i}-\frac{1}{4\pi}\,\bmag\otimes\bmag\right)= (A_{\rm p}\,\rho_{\rm p}+A_{\rm CI\hspace{-.1em}I}\,\rho_{\rm CI\hspace{-.1em}I})\,\rho_{\rm n}\,(\bvel_{\rm n}-\bvel_{\rm i}),\\
&& \frac{\partial e_{\rm n}}{\partial t}+\bnabla\cdot\{(e_{\rm n}+p_{\rm n})\,\bvel_{\rm n}\}=
\bnabla\kappa\bnabla T_{\rm n}-\rho_{\rm n}\,\mathcal{L} + \Big( \frac{A_{\rm p}\,m_{\rm p}\,\rho_{\rm p}}{m_{\rm n}+m_{\rm p}}+\frac{A_{\rm CI\hspace{-.1em}I}\,m_{\rm CI\hspace{-.1em}I}\,\rho_{\rm CI\hspace{-.1em}I}}{m_{\rm n}+m_{\rm CI\hspace{-.1em}I}} \Big)\,\rho_{\rm n}\,(\bvel_{\rm i}-\bvel_{\rm n})^{2},\label{E5}\\
&& \frac{\partial\bmag}{\partial t}=\bnabla\times(\bvel_{\rm i}\times\bmag),
\end{eqnarray}
where $e=p/(\gamma-1)+\rho\,v^{2}/2$ is the total energy per volume and $\vec{\nabla}=(\partial/\partial x,\,\partial/\partial y,\,0)$ is the two-dimensional spatial derivative operator.
The subscripts $n$ and $i$ denote the neutral and charged components, respectively.
In the neutral component, we take into account hydrogen and helium atoms.
In the charged component, we take into account hydrogen, helium and carbon ions.
The ionized hydrogens and heliums are described with the subscript $\rm p$, and the ionized carbons are described with the subscript $\rm CI\hspace{-.1em}I$ ($\rho_{\rm i}=\rho_{\rm p}+\rho_{\rm CI\hspace{-.1em}I}$).
We assume that the density of the ionized carbon is proportional to the density of the neutral gas $\rho_{\rm CI\hspace{-.1em}I}=\mathcal{A}_{\rm C} \rho_{\rm n} m_{\rm CI\hspace{-.1em}I}/m_{\rm n}$, where we use $\mathcal{A}_{\rm C}=3\times 10^{-4}$.
In addition, we impose the ideal gas equation of state $p_{\rm n}=k_{\rm B}\rho_{\rm n}T/m_{\rm n}$, and we use the isothermal approximation between the neutral gas and the ionized gas $p_{\rm i} = p_{\rm n}\,m_{\rm n}\,\rho_{\rm i}/m_{\rm i}/\rho_{\rm n}$, where the mean charged particle mass is determined by $m_{\rm i}=(\rho_{\rm p}+\rho_{\rm CI\hspace{-.1em}I})/(n_{\rm p}+n_{\rm CI\hspace{-.1em}I})$.
We use the specific heat ratio $\gamma=5/3$ and the mean particle masses $m_{\rm n}=m_{\rm p}=1.27\,m_{\rm proton}$ (i.e., we assume for convenience that 91\% of neutrals and light ions are, respectively, hydrogens and hydrogen ions, and that the remaining 9\% of neutrals and light ions are, respectively, heliums and helium ions).
The drag coefficients $A_{\rm p}$ and $A_{\rm CI\hspace{-.1em}I}$, the recombination coefficient $\alpha$, the ionization coefficient $\xi$, and the thermal conductivity $\kappa$ are chosen to simulate the typical ISM whose properties are listed in Table 1.

\subsection{Cooling Function and Its Characteristics}
We take the following simplified net cooling function:
\begin{eqnarray}
\rho_{\rm n}\,\mathcal{L} &=& n_{\rm n}\,(-\Gamma +n_{\rm n}\,\Lambda)\,\mbox{ erg cm}^{-3}\mbox{ s}^{-1}, \label{CF}\\
\Gamma &=& 2\times 10^{-26},\\
\frac{\Lambda}{\Gamma} &=& 1.0\times 10^{7}\,\exp\left( \frac{-118400}{T+1000} \right) + 1.4\times 10^{-2}\,\sqrt{T}\,\exp\left( \frac{-92}{T} \right).
\end{eqnarray}
This is obtained by fitting to the various heating ($\Gamma$) and cooling ($\Lambda$) processes considered by Koyama \& Inutsuka (2000).
The resulting thermal-equilibrium state defined by $\mathcal{L}(n_{\rm n},T)=0$ is shown in Figure \ref{f1} as a thick solid line.
We also show the contour of the timescale of the cooling and heating $t_{\rm cool}=k_{\rm B}T/\{ m_{\rm n}\mathcal{L}\,(\gamma-1)\}$ as broken lines.

Isochorically cooling gas and isobarically contracting gas, which can be generated by shock compression, are thermally unstable, if these gases satisfy the Balbus criterion $[\partial (\mathcal{L}/T)/\partial T]_{p}<0$ (Balbus 1995)\footnote{The linear stability analysis of isochorically cooling gas was done by Schwarz, McCray, \& Stein (1972) and that of isobarically contracting gas was done in Appendix B of Koyama \& Inustuka (2000).}.
In Figure \ref{f4}, we show the unstable region as a dotted region.

\section{Numerics}
\subsection{Numerical Schemes}
We use the operator-splitting technique for solving the basic equations which are split into four parts: (1) the ideal HD part for neutral gas, (2) the ideal MHD part for ionized gas, (3) the cooling, heating and thermal conduction part, and (4) the ionization, recombination, and frictional forces/heating part. 

Methods that solve parts 1, 2, and 3 are well established.
We solve the ideal HD part by employing the second-order Godunov method (van Leer 1979).
The ideal MHD parts are solved by employing the second-order method developed by Sano, Inutsuka \& Miyama (1999), in which the HD parts with the magnetic pressure term are solved using second-order Godunov method, and the magnetic tension term and the induction equation are solved using the method of characteristics for Alfv\'en waves (Stone \& Norman 1992) with the constrained transport algorithm (Evans \& Hawley 1988).
We use the explicit time integration for the cooling, heating and thermal conduction, which has second-order accuracy in time and space.

In the parts 1 and 2, equations are solved in conservative fashion.
Thus, the total mass, momentum, and energy of neutral gas and the total mass and momentum of ionized gas conserve exactly in the absence of source terms.
The internal energy of ions is determined by equating temperatures of neutral and ionized gases.
This approximation is reliable when the ionization degree is small and the relaxation time is short.
In HI regions considered in this paper, both conditions are fulfilled.
It is known that, when the plasma $\beta$ is small ($\lesssim 0.01$), the conservative MHD scheme that uses total energy ($E=e+B^{2}/8\pi$) as a basic variable to update cannot describe thermal energy accurately, because thermal energy should be calculated by subtracting magnetic energy ($B^{2}/8\pi$) and kinetic energy ($\rho\,v^{2}/2$) from the total energy ($E$).
In the simulation of weakly ionized plasma employing a two-fluid conservative method, we would encounter this problem even when the $\beta$ of the whole medium is not small, since the plasma $\beta$ of ionized gas is always lower by a factor of the ionization degree.
However, if we assume the temperatures of neutral and ionized gases are the same, we can perform the simulation without such difficulty, as in this paper.

In part 4, we use the method presented in the next section which utilizes exact solutions of split equations and is unconditionally stable.
Thus, part 4 does not restrict the time step, and the time step in our code as a whole is determined so that the following criteria are satisfied everywhere: the CFL condition for the sound wave of neutral gas ($\Delta t_{\rm n}<0.5\,\Delta\,x/(c_{\rm s}+|v_{\rm n}|)$), the CFL condition for the fast wave of ionized gas ($\Delta t_{\rm i}<0.5\,\Delta\,x/(c_{\rm fast}+|v_{\rm i}|)$), the condition for avoidance over-cooling/heating ($\Delta t_{\rm cool}<0.2\,p_{\rm n}/[\,\rho_{\rm n}\,|\mathcal{L}|\,(\gamma-1)\,]$), and the stability condition for thermal conduction ($\Delta t_{\rm cond}<0.5\,(\Delta x)^{2}\,n_{\rm n}\,k_{\rm B}/[\,(\gamma-1)\,\kappa\,]$).

\subsection{Source Term Solver}
\subsubsection{Exact Maps}
In this section we present a scheme for solving the source part in the basic equations (\ref{E1})-(\ref{E5}).
Let us consider differential equations
\begin{eqnarray}
\frac{\partial f}{\partial t}&=&g_{1}(f)+g_{2}(f),\\
\frac{\partial \tilde{f}}{\partial t}&=&g_{2}(\tilde{f}),\\
\frac{\partial \hat{f}}{\partial t}&=&g_{1}(\hat{f}),
\end{eqnarray}
where $g$ is an operator which can include the spatial derivatives of $f$.
If we introduce a map $G(\Delta t, f)$ that gives a solution of a differential equation at $t+\Delta t$ with second-order accuracy in time, 
\begin{eqnarray}
f(t+\Delta t)&=&G_{12}(\Delta t, f(t))+\mathcal{O}(\Delta t^{3}),\\
\tilde{f}(t+\Delta t)&=&G_{1}(\Delta t, \tilde{f}(t))+\mathcal{O}(\Delta t^{3}),\\
\hat{f}(t+\Delta t)&=&G_{2}(\Delta t, \hat{f}(t))+\mathcal{O}(\Delta t^{3}),
\end{eqnarray}
then it is known that 
the operator $G_{12}$ can be approximated by 
the combination of $G_{1}$ and $G_{2}$,
\begin{equation}\label{ID}
G_{12}(\Delta t,f)=G_{2}^{\Delta t/2}\,G_{1}^{\Delta t}\,G_{2}^{\Delta t/2}\,f+\mathcal{O}(\Delta t^{3}).
\end{equation}
where we have introduced a convention $G(\Delta t_{1},H(\Delta t_{2},f))=G^{\Delta t_{1}}\,H^{\Delta t_{2}}\,f$.
In the case of basic equations (\ref{E1})-(\ref{E5}), the operator $g_{1}$ corresponds to HD, MHD, cooling, heating, and thermal conduction operators, which are also solved by using further operator splitting explained in \S3.1 , and the function $g_{2}$ corresponds to other source terms.
The approximation from equation (\ref{ID}) remains valid,  as long as the map $G_{2}$ is more accurate than the second-order. 
In our case, the split source equations are the following,
\begin{eqnarray}
\frac{d\tilde{n}_{\rm n}}{dt}&=&-\xi\,\tilde{n}_{\rm n},\\
\frac{d\tilde{n}_{\rm p}}{dt}&=&\xi\,\tilde{n}_{\rm n},\\
\frac{d\hat{n}_{\rm n}}{dt}&=&\alpha\,\hat{n}_{\rm p}^{2},\\
\frac{d\hat{n}_{\rm p}}{dt}&=&-\alpha\,\hat{n}_{\rm p}^{2},\\
\frac{d\hat{\bvel}_{\rm n}}{dt}&=&A_{1}\,(\hat{\bvel}_{\rm i}-\hat{\bvel}_{\rm n})\\
\frac{d\hat{\bvel}_{\rm i}}{dt}&=&A_{2}\,(\hat{\bvel}_{\rm n}-\hat{\bvel}_{\rm i})\\
\frac{d\hat{p}_{\rm i}}{dt}&=& (\gamma-1)\,A_{3}\,(\hat{\bvel}_{\rm i}-\hat{\bvel}_{\rm n})^{2}.
\end{eqnarray}
We can derive the following exact map ($t\rightarrow t+\Delta t$): 
\begin{eqnarray}
\tilde{n}_{\rm n}(t+\Delta t) &=& \tilde{n}_{\rm n}(t)\,\exp^{-\xi\,\Delta\,t},\\
\tilde{n}_{\rm i}(t+\Delta t) &=& \tilde{n}_{\rm i}(t)+\tilde{n}_{\rm n}(t)\,(1-\exp^{-\xi\,\Delta\,t}),\\
\hat{n}_{\rm n}(t+\Delta t) &=& \frac{\hat{n}_{\rm n}(t)+\alpha\,\Delta t\,\hat{n}_{\rm p}(t)\,\{\hat{n}_{\rm n}(t)+\hat{n}_{\rm p}(t)\}}{1+\alpha\,\Delta t\,\hat{n}_{\rm p}(t)},\\
\hat{n}_{\rm i}(t+\Delta t) &=& \frac{\hat{n}_{\rm p}(t)}{1+\alpha\,\Delta t\,\hat{n}_{\rm p}(t)},\\
\hat{\bvel}_{\rm n}(t+\Delta t) &=& \frac{1}{A_{1}+A_{2}}\{\,\hat{\bvel}_{\rm i}(t)\,A_{1}\,( 1-\exp^{-(A_{1}+A_{2})\,\Delta t} ) \nonumber\\&& +\hat{\bvel}_{\rm n}(t)\,( A_{\rm 2}+A_{\rm 1}\,\exp^{-(A_{1}+A_{2})\,\Delta t} ) \},\label{STS1}\\
\hat{\bvel}_{\rm i}(t+\Delta t) &=& \frac{1}{A_{1}+A_{2}}\{\,\hat{\bvel}_{\rm n}(t)\,A_{2}\,( 1-\exp^{-(A_{1}+A_{2})\,\Delta t} ) \nonumber\\&& +\hat{\bvel}_{\rm i}(t)\,( A_{\rm 1}+A_{\rm 2}\,\exp^{-(A_{1}+A_{2})\,\Delta t} ) \},\label{STS2}\\
\hat{p}_{\rm n}(t+\Delta t) &=& \hat{p}(t)+\frac{(\gamma-1)\,A_{3}\,\{\hat{\bvel}_{i}(t)-\hat{\bvel}_{n}(t)\}^{2}}{2\,(A_{1}+A_{2})}\,\{ 1-\exp^{-2\,(A_{1}+A_{2})\,\Delta t} \},
\end{eqnarray}
where $A_{\rm 1}\equiv A_{\rm p}\,\rho_{\rm p}+A_{\rm CI\hspace{-.1em}I}\,\rho_{\rm CI\hspace{-.1em}I}$, 
$A_{\rm 2}\equiv (A_{\rm p}\,\rho_{\rm p}+A_{\rm CI\hspace{-.1em}I}\,\rho_{\rm CI\hspace{-.1em}I})\,\rho_{\rm n}/\rho_{\rm i}$, and 
$A_{3}\equiv \rho_{\rm n}\,\{(A_{\rm p}\,m_{\rm p}\,\rho_{\rm p})/(m_{\rm n}+m_{\rm p})+(A_{\rm CI\hspace{-.1em}I}\,m_{\rm CI\hspace{-.1em}I}\,\rho_{\rm CI\hspace{-.1em}I})/(m_{\rm n}+m_{\rm CI\hspace{-.1em}I})\}$, 
and we assume that the coefficient $A_{\rm p}$ 
(a function of temperature)
is constant during time step $\Delta t$.
Even in the case where we do not use the equal temperature approximation between neutral and ionized gases, we can find the exact maps for $p_{\rm n}$ and $p_{\rm i}$
that were used in Inoue, Inutsuka \& Koyama (2007).

Since variables are evolved according to the exact solutions, this part is unconditionally stable and does not limit the time step.
Furthermore, if $G_{2}$ is a exact map, 
the time evolution ($t \rightarrow t+\Sigma_{i=1}^{n}\Delta t$)
can be expressed as 
\begin{eqnarray}
f(t+\sum_{i=1}^{n}\,\Delta t_{i})&=&G_{2}^{\Delta t_{n}/2}\,G_{1}^{\Delta t_{n}}\,G_{2}^{\Delta t_{n}/2}\,
G_{2}^{\Delta t_{n-1}/2}\,G_{1}^{\Delta t_{n-1}}\,G_{2}^{\Delta t_{n-1}/2}\,\cdots\,\nonumber\\&&
G_{2}^{\Delta t_{1}/2}\,G_{1}^{\Delta t_{1}}\,G_{2}^{\Delta t_{1}/2}\,f(t)+\mathcal{O}(\Delta t^{3})\label{SC1}\\
&=& G_{2}^{\Delta t_{n}/2}\,G_{1}^{\Delta t_{n}}\,G_{2}^{(\Delta t_{n}+\Delta t_{n-1})/2}\,
G_{1}^{\Delta t_{n-1}}\,G_{2}^{(\Delta t_{n-1}+\Delta t_{n-2})/2}\cdots\nonumber\\&&
G_{2}^{(\Delta t_{2}+\Delta t_{1})/2}\,G_{1}^{\Delta t_{1}}\,G_{2}^{\Delta t_{1}/2}\,f(t)+\mathcal{O}(\Delta t^{3}).\label{SC2}
\end{eqnarray}
In equation (\ref{SC1}), there are $3\,n$ operators, while they become $2\,n+1$ operators in equation (\ref{SC2}).
Thus, by using the exact map, we can save computation time without loss of accuracy.

\subsubsection{Tests for Alfv\'en Waves with Plasma Drift}
\begin{figure}[t]
\epsscale{0.8}
\plotone{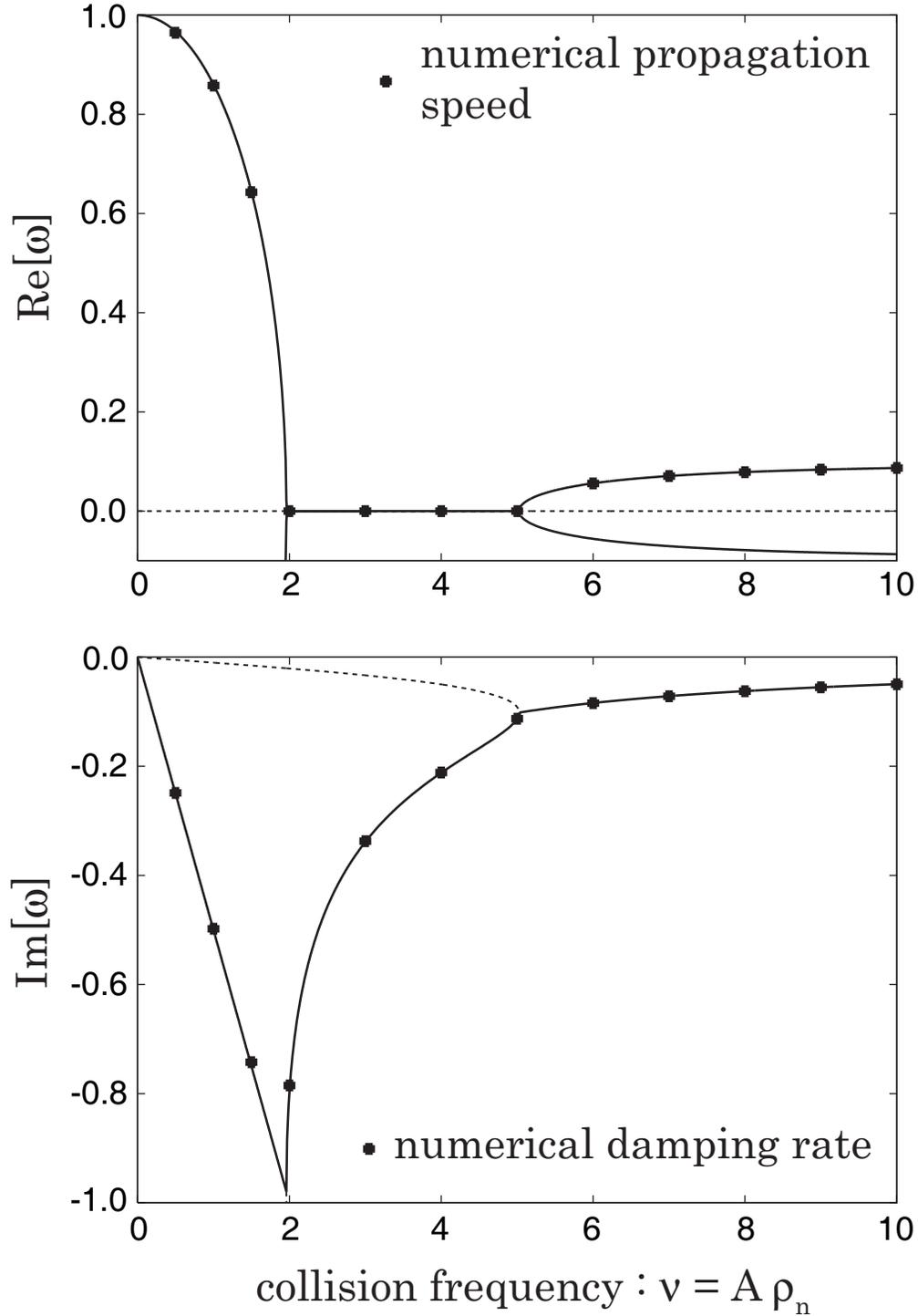}
\caption{
Real (\textit{top}) and imaginary (\textit{bottom}) parts of $\omega$ against the collision frequency $\nu$ in the case $B_{x}=1.0,\,\rho_{\rm n}=1.0\times 10^{2},\,\rho_{\rm i}=1.0,$ and $k=1.0$.
The solid lines are the branches of the Alfv\'en waves that propagate in the weak ($0\le\nu\le 1.96$) and strong ($\nu\ge5.03$) coupling regimes.
The dashed lines show the other mode, i.e., a purely damping mode.
Propagation speeds (numerical Re$[\omega]$) and damping rates (numerical Im$[\omega]$) obtained from test runs with cell number $m=128$ are plotted as points in the cases that $\nu=0.5,\,1.0,\,1.5,\,2.0,\,3.0,\,4.0,\,5.0,\,6.0,\,7.0,\,8.0,\,9.0,$ and $10.0$.
}
\label{f1}
\end{figure}

In order to confirm the capability of our method, 
we perform test calculations with various collision frequencies.
We also verify our method by a convergence test. 
The basic equations used in these tests are
\begin{equation}\label{KP1}
\rho_{\rm n}\,\frac{\partial v_{{\rm n},y}}{\partial t}=A\,\rho_{\rm n}\,\rho_{\rm i}\,(v_{{\rm i},y}-v_{{\rm n},y}),
\end{equation}
\begin{equation}
\rho_{\rm i}\,\frac{\partial v_{{\rm i},y}}{\partial t}=B_{x}\,\frac{\partial B_{y}}{\partial x}+A\,\rho_{\rm n}\,\rho_{\rm i}\,(v_{{\rm n},y}-v_{{\rm i},y}),
\end{equation}
\begin{equation}\label{KP3}
\frac{\partial B_{y}}{\partial t}=B_{x}\,\frac{\partial v_{{\rm i},y}}{\partial x}.
\end{equation}
If we assume $\rho_{\rm n},\,\rho_{\rm n},\,B_{\rm x}$ and $A$ are constants, these equations become linear differential equations 
that describe the evolution of Alfv\'en waves in a uniform partially ionized medium.
Assuming the solutions to be $\propto \exp i(k\,x-\omega\,t)$, the characteristic equation can be written as (Kulsrud \& Pearce 1969)
\begin{equation}\label{CHA}
\omega\,(\omega^{2}-c_{\rm A}^{2}\,k^{2})+i\,\nu\,\left\{ (1+\epsilon)\,\omega^{2}-c_{\rm A}^{2}\,k^{2}\,\epsilon \right\}=0,
\end{equation}
where $\nu=A\,\rho_{\rm n}$ is the collision frequency of a charged particle in a sea of neutrals, $\epsilon=\rho_{\rm i}/\rho_{\rm n}$ is the ionization degree, and $c_{\rm A}=B_{x}/\rho_{\rm i}^{1/2}$ is the Alfv\'en velocity in the weak-coupling limit.
This characteristic equation contains three modes.
In Figure \ref{f1}, we plot the real (top) and imaginary (bottom) parts of $\omega$ against the collision frequency $\nu$.
The other parameters are chosen such that $B_{x}=1.0,\,\rho_{\rm n}=1.0\times 10^{2},\,\rho_{\rm i}=1.0,$ and $k=1.0$, which are used throughout in test calculations.
The solid lines are the branches of the Alfv\'en waves that propagate in the weak ($0\le\nu\le 1.96$) and strong ($\nu\ge5.03$) coupling regimes.
The dashed lines show the other mode, i.e., a purely damping mode.
In the weak-coupling limit ($\nu \rightarrow 0$), the phase velocity of the Alfven wave is given by $B_{x}/\rho_{\rm i}^{1/2}=1.0$, and in the strong-coupling limit ($\nu \rightarrow \infty$), it is given by $B_{x}/(\rho_{\rm i}+\rho_{\rm n})^{1/2}=9.95\times 10^{-2}$.

To demonstrate the propagation of Alfven waves, we set the initial condition by eigenfunctions of three modes,
\begin{equation}
B_{y}(x)={\rm Re}\left[ e^{i\,x} \right],
\end{equation}
\begin{equation}
v_{{\rm i},y}(x)=-{\rm Re}\left[\omega\,B_{y}(x)\right],
\end{equation}
\begin{equation}
v_{{\rm n},y}(x)=v_{{\rm i},y}(x)+{\rm Re}\left[\frac{i}{\nu}\,(\omega^{2}-1 )\,B_{y}(x)\right].
\end{equation}
We take the $\omega$ that is the solution of the characteristic equation (\ref{CHA}) and is on the branch of the thick solid line in Figure \ref{f1}.
In the test code, we use an operator-splitting technique based on equation (\ref{SC2}) in which we use the MOC scheme (Stone \& Norman 1992) and the exact map of the friction part based on equations (\ref{STS1}) and (\ref{STS2}).
The former one solves the propagation of Alfv\'en waves in the absence of the friction term, and the latter one solves the friction force term.
We use a periodic domain that covers the $x\in [0.0,2\pi]$ region.
The time step $\Delta t$ is determined from the CFL condition for the Alfv\'en wave (in this case, $\Delta t=\pi/m$, where $m$ is the cell number used in a run).

\begin{figure}[t]
\epsscale{1.}
\plotone{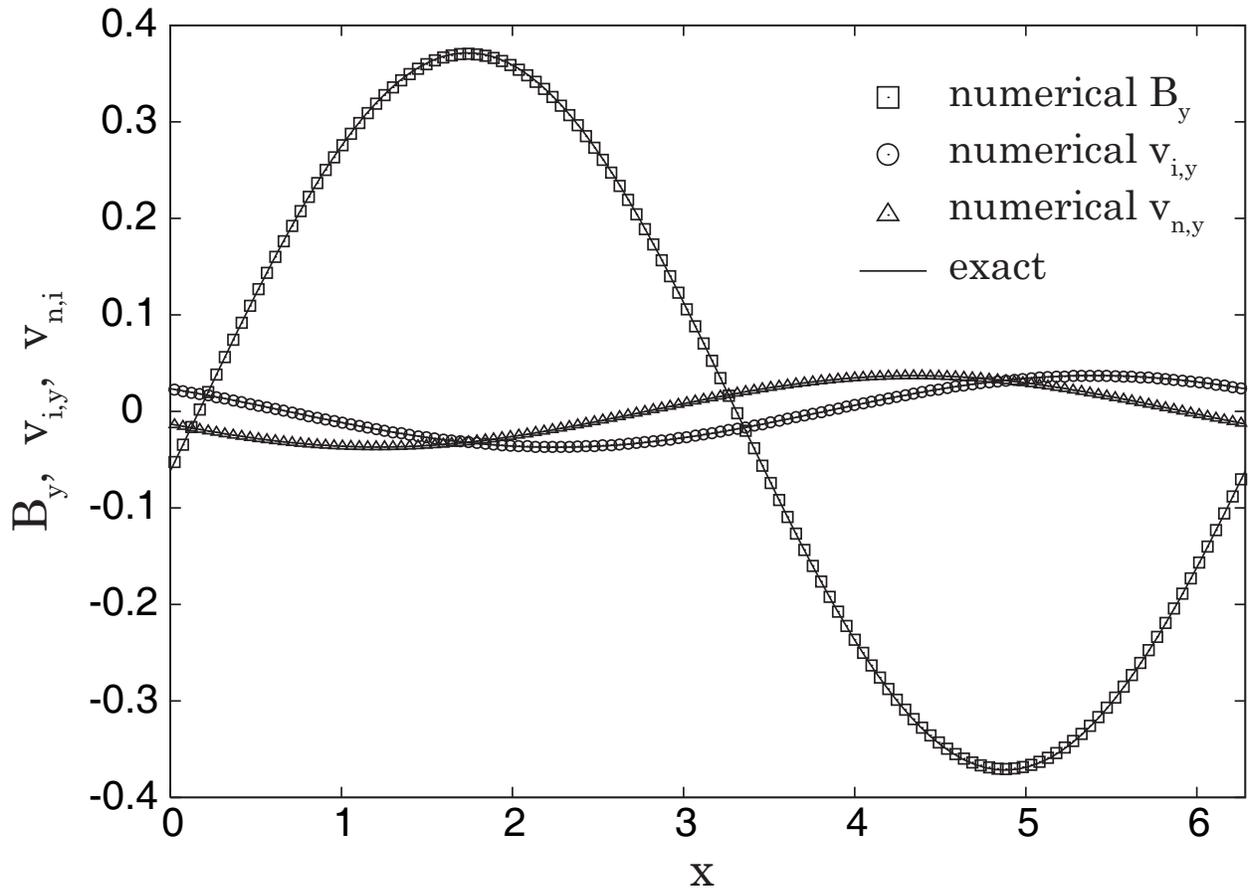}
\caption{
Numerical (\textit{points}) and exact (\textit{solid lines}) solutions at $t=20.0$.
The number of cells is $m=128$ in the calculations.
}
\label{f2}
\end{figure}

\begin{figure}[t]
\epsscale{1.}
\plotone{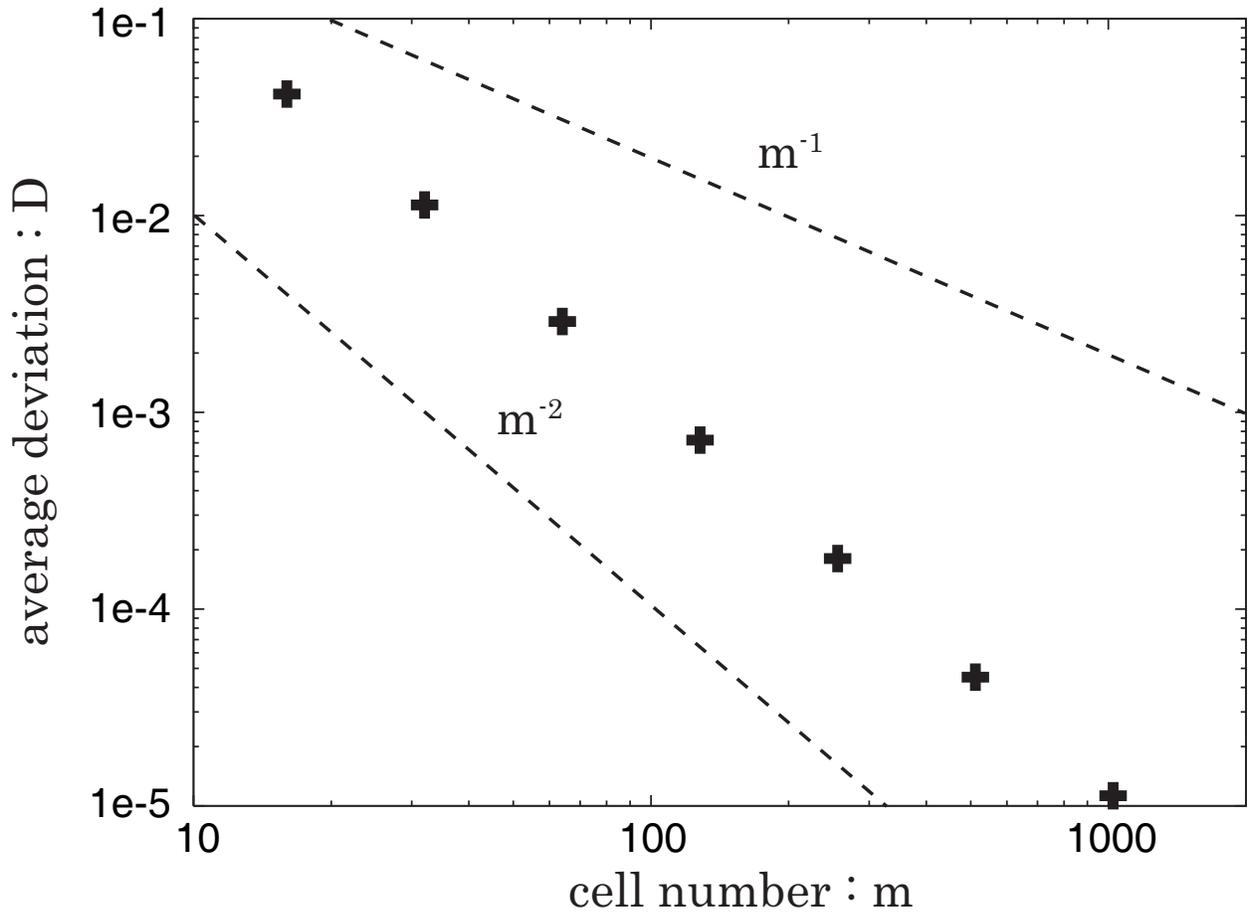}
\caption{
Average deviation from exact solution at $t=20.0$ as a function of cell number $m$.
We also show reference lines $D\propto m^{-1} and m^{-2}$ as dashed lines.
}
\label{f3}
\end{figure}

The second test is a convergence test for the case $\nu=10.0$.
Calculations are performed from $t=0.0$ to $20.0$, while the Alfv\'en waves propagate about 0.28 wavelengths and are damped by a factor $0.37$.
Since basic equations are linear differential equations, we can make direct comparison between numerical and exact solutions.
In Figure \ref{f2} we show both the numerical solution with $m=128$ and the exact solution at $t=20.0$.
The numerical solution agrees well with the exact solution.
In Figure \ref{f3} we show the plot of average deviation from the exact solution at $t=20.0$,
\begin{equation}
D(m)=\frac{1}{m}\,\sum_{i=1}^{m}\left|B_{y,i}-B_{y}(x_{i})\right|_{t=20.0},
\end{equation}
as a function of cell number $m$ used in calculations, where $B_{y,i}$ is the magnetic field at the $i$-th cell center calculated numerically, and $B_{y}(x)$ is the exact solution.
This figure clearly shows that the second-order accuracy in time and space is achieved in our scheme based on the exact map and the transformation corresponding to equation (\ref{SC2}).

In order to confirm the capability of our method from the weak-coupling regime to the strong-coupling regime, we examine the propagation of Alfv\'en waves in a partially ionized medium with various collision frequencies.
The propagation speed of the numerical wave can be determined by tracing the position where $B_{y}$ is maximum.
We measure the propagation speed and calculate numerical ${\rm Re}[\omega]$ by using the time when the position of maximum $B_{y}$ crosses $x=2\pi$, except the cases of $1.96\le\nu\le5.03$ in which the propagation speed is measured at the time $t=20.0$.
The damping rate (numerical ${\rm Im}[\omega]$) is measured by using the time when $\int_{0}^{2\pi}|B_{y}(x,t)|dx/\int_{0}^{2\pi}|B_{y}(x,0)|dx$ becomes smaller than $1/e$.
In Figure \ref{f1} we plot the numerically obtained $\omega$ by $m=128$ calculations as points.
This figure clearly shows that our method can solve the two-fluid system in all the cases from the weak-coupling regime to strong-coupling regime.

\subsection{Initial Settings}
In order to study the generation of clouds as a result of TI, we consider the situation in which supersonic WNM flows collide with supersonic velocity, which will supply thermally unstable gas in the shocked slab.
A rectangle computational domain is used whose side lengths are $30.0$ pc for the $x$-direction ($x=-15.0$ to $15.0$ pc) and $2.0$ pc for the $y$-direction ($y=-1.0$ to $1.0$ pc).
Two opposing WNM flows with density fluctuations initially collide head-on at the $x=0.0$ interface.
Periodic boundary condition is imposed on boundaries at $y=-1.0$ and $1.0$ pc, and WNM flows without fluctuations are constantly set at $x=\pm15.0$ pc boundaries.
Note that the density fluctuations are added only initially, and their properties are described below. 
We use nonuniform mesh.
The region of $x=-1.0$ to $1.0$ and $y=-1.0$ to $1.0$ pc is divided into $2048\times 2048$ cells, and the regions of $x=-15.0\,(15.0)$ to $-1.0\,(1.0)$ pc and $y=-1.0$ to $1.0$ pc are divided into $512\times 2048$ cells.
Thus, the spatial resolutions are $\Delta x= 9.8 \times 10^{-4}$ pc in the higher resolution region, $\Delta x= 2.7 \times 10^{-2}$ pc in the lower resolution region, and $\Delta y= 9.8 \times 10^{-4}$ pc.
In the higher resolution region, the Field condition (Koyama \& Inutsuka 2004) is satisfied.

We choose the density and pressure of the unperturbed WNM as $\rho_{\rm w}=0.57\,m_{\rm n}$ and $p_{\rm w}=3.5\times10^{3}\,k_{\rm B}$, which are the thermal equilibrium values of the cooling function (eq. [\ref{CF}]).
In order to excite TI, we add fluctuations in the initial density of the WNM as
\begin{equation}
\log \rho_{n,\rm ini} = \left( 1+\mathcal{A}\,\sum_{l_{x}=2}^{6}\sum_{k_{y}=1}^{5}\,\,\sin\left[\frac{2\pi\,x}{l_{x}}+\theta(l_{x})\,\right]\,\sin\left[\frac{2\pi\,k_{y}\,y}{L_{y}}+\theta(l_{y})\,\right] \right)\,\log \rho_{\rm w},
\end{equation}
where $\rho_{n, \rm eq}$ is the neutral gas density of the unstable equilibrium, $\theta(l)$ is a random phase, and we use $\mathcal{A}=0.1$.
By doing so, fluctuations whose scales are around $\lambda = 1$ pc (the most unstable scale of TI) are generated in the shocked layer as a result of the WNM flows collision.
The minimum and maximum density of the perturbed initial condition are $\rho_{\rm n,min}=0.37\,m_{\rm n}$ and $\rho_{\rm n,max}=0.93\,m_{\rm n}$.
The initial weakly ionized gas is set as an ionization equilibrium.

The initial magnetic field strength is $2.0\,\mu$G, which might be a typical value in the WNM.
The initial orientation of the magnetic field is perpendicular to the initial inflow velocity, so that the effect of magnetic field becomes the strongest.
The effect of the orientation of the magnetic field is discussed in \S 4.
In order to contrast the effect of magnetic field, we also performed the same calculation except in the absence of magnetic field.

We set the initial converging velocity at $20.0$ km s$^{-1}$, with which the Mach number is $M\equiv |v_{x}|/c_{\rm fast}=2.1$ for the unperturbed gas of the magnetized case and $M\equiv |v_{x}|/c_{\rm s}=2.4$ for that of the unmagnetized case.
We also performed several calculations decreasing the initial velocity down to $M=1$ both with and without magnetic field, and confirmed that the results shown below were not qualitatively changed.
Note that as shown in Hennebelle \& Perault (1999), if the speeds of the converging flows are subsonic, thermal condensation due to TI does not take place.

\section{Results}
The evolutions of the shocked WNM are very different in magnetized and unmagnetized cases:
\begin{itemize}
\item \textit{Unmagnetized case}.
\end{itemize}

\begin{figure}[t]
\epsscale{1.}
\plotone{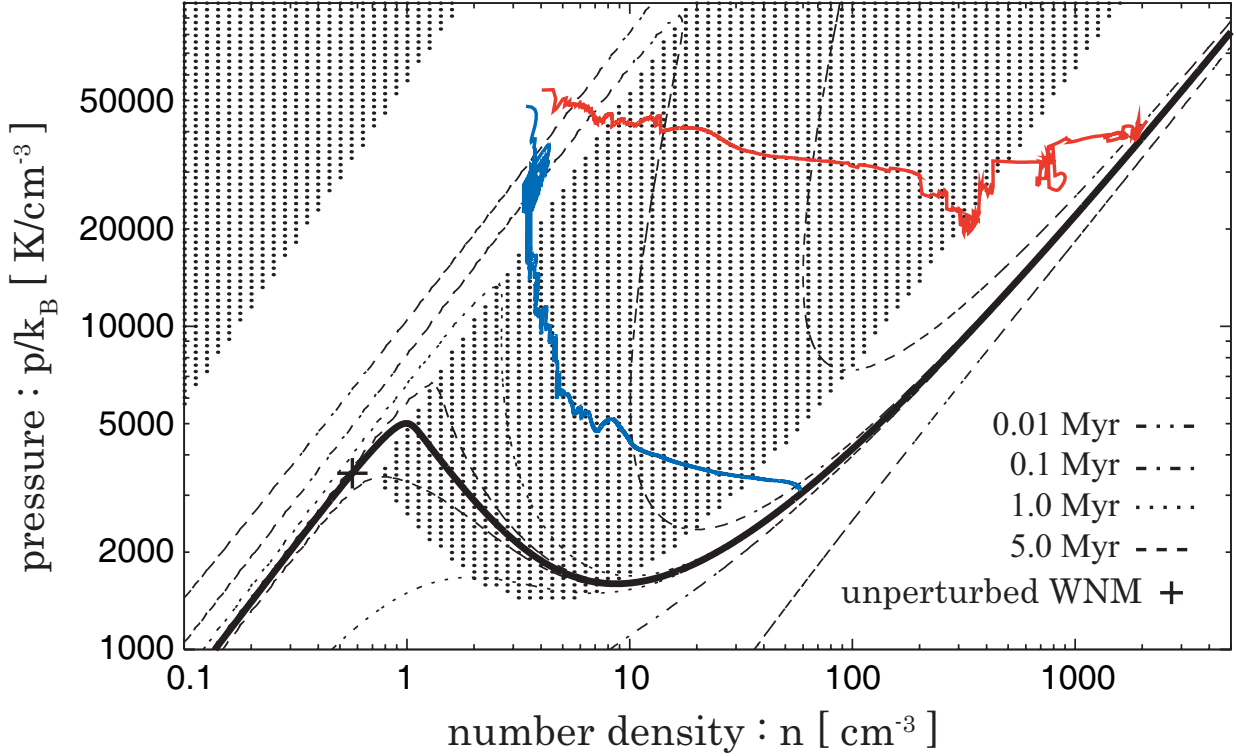}
\caption{
Thermal-equilibrium state of the net cooling function (\textit{solid line}).
Contours of the cooling and heating timescales, which correspond to $5.0,\,1.0,\,0.1,$ and $0.01$ Myr, are shown as broken lines.
Dotted region shows thermally unstable region, in which the Balbus criterion is satisfied.
Red and blue lines are the evolutionary tracks of the highest density gas.
Red line shows the unmagnetized case, which reaches the stable CNM phase at $t=0.5$ Myr.
Blue line shows the magnetized case, which reaches the stable CNM phase at $t=1.6$ Myr.
}
\label{f4}
\end{figure}

At first, as indicated in Koyama \& Inutsuka (2000), shock-compressed gas gradually contract almost isobarically, until the gas satisfies the Balbus criterion.
Once the gas becomes thermally unstable, overdensity perturbations begin to condense exponentially in time, until these regions reach a stable CNM phase.
In Figure \ref{f4} we show the evolutionary track of the highest density gas in the computational domain as a red line until it reaches the stable CNM phase, which is at $t=0.5$ Myr.

\begin{figure}[t]
\epsscale{1.}
\plotone{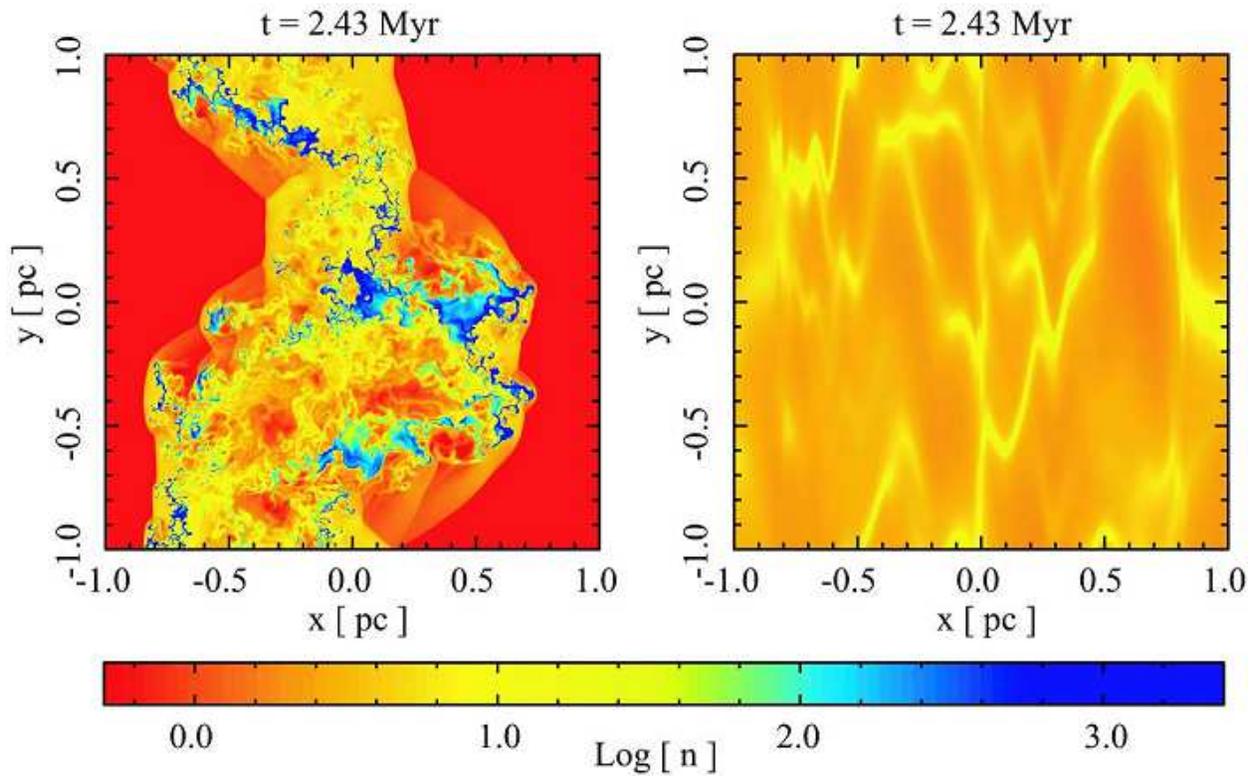}
\caption{
Density distribution of unmagnetized (\textit{top left}) and magnetized (\textit{top right}) cases.
Density distributions are depicted with the same color scale (\textit{bottom}).
}
\label{f5}
\end{figure}

The density distribution at time $t=2.43$ Myr is shown in the left panel of Figure \ref{f5}.
Inside the shocked slab, CNM cloudlets are generated, and the interwoven medium composed of the cloudlets, unstable gas, and WNM is formed.
In the core of the generated cloudlets, number density reaches $n\gtrsim 1\times10^{3}$ cm$^{-3}$.
Since, these small cloudlets move randomly with supersonic velocity dispersion, as first pointed out by Koyama \& Inutsuka (2002), the shocked slab has complex geometry (see also Heitsch et al. 2005).

Despite the fact that the shocked gas is cooled until it reaches the CNM phase, the thermal pressure of the CNM is maintained at a high value ($p/k_{\rm B}\gtrsim 1\times10^{4}$ K cm$^{-3}$).
This is because the CNM and surrounding thermally unstable gas slab are bound by the ram pressure of the converging flows.
When the cooling rate in the slab balances the thermalization rate of the shocks, outward propagating shocks become quasi-standing shocks.
This transition takes place at $t\simeq 0.5$ Myr in this calculation, which is roughly simultaneous with the time of the CNM formation.
Thus, as long as the converging flows continue, high-density and high-pressure conditions in the CNM are held, and the mass of the CNM continues to increase owing to the accretion of the ``fresh" unstable gas which is continuously supplied by the quasi-standing shocks.
Such a CNM would evolve into a molecular cloud.
Similar results are also reported in the papers introduced in \S 1.

\begin{itemize}
\item \textit{Magnetized case}.
\end{itemize}

At first, shock-compressed gas isochorically cools, decreasing its thermal pressure, until the gas satisfies the Balbus criterion.
Once the gas becomes thermally unstable, overdensity perturbations begin to condense.
During the growth of the TI (condensation), the thermal pressure continues to decrease due to the cooling.
Since the growth timescale of the most unstable scale of TI is equal to the timescale of the cooling (Field 1965), by the time condensing regions reach a stable CNM phase, the thermal pressure has fallen $p/k_{\rm B}\simeq 2\times10^{3}$ K cm$^{-3}$, which is on the order of the average pressure of the ISM.
In Figure \ref{f4} we show the evolutionary track of the highest density gas as a blue line until it reaches the stable CNM phase, which is at $t=1.6$ Myr.
This evolution is quite different from the case of the unmagnetized result.
This is because magnetic pressure in the shocked slab can sustain the ram pressure of the converging flows, and the thermal pressure does not have to be high in order to oppose the ram pressure.
Thus, in contrast to the unmagnetized case, the shocks never become standing shocks and the width of the post shock layer increases almost linearly with time.

\begin{figure}[t]
\epsscale{1.}
\plotone{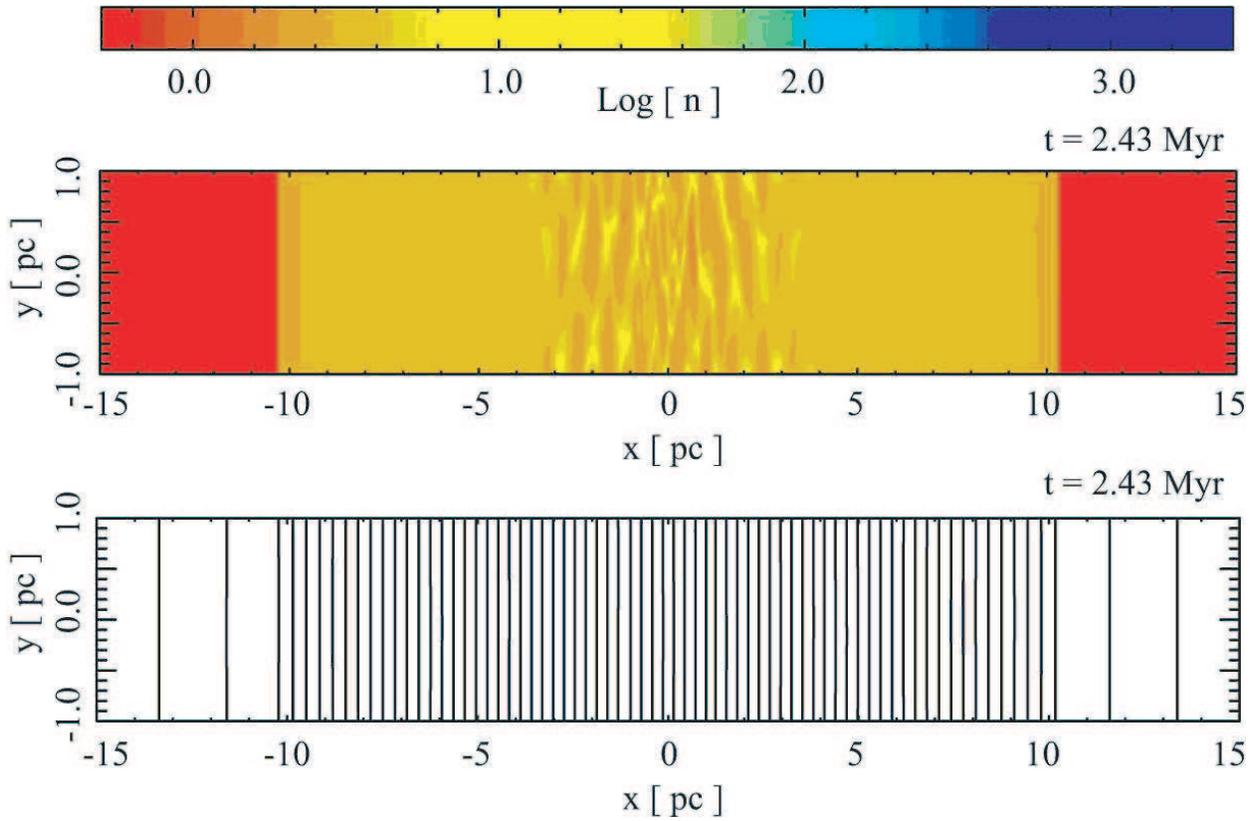}
\caption{
Density distribution (\textit{top}) and magnetic field lines (\textit{bottom}) of the whole simulation domain at the same time depicted in the right panel of Figure \ref{f5}.
Density distributions are depicted with the same color scale to Figure \ref{f5}.
Note that the y-scale is expanded in these figures.
}
\label{f6}
\end{figure}

The density distribution at time $t=2.43$ Myr is shown in the right panel of Figure \ref{f5}.
Because the decrease of the thermal pressure leads to longer cooling times (see contours of cooling time in Figure \ref{f4}), the nonlinear growth of the TI in the magnetized case needs a longer time than the unmagnetized case.
At this time, most of dense clumps have reached the stable CNM phase in the region shown in the figure.
We also show the density distribution (Fig. 6, \textit{top}) and magnetic field lines (Fig. 6, \textit{bottom}) of the whole simulation domain in Figure \ref{f6} at the same time depicted in the right panel of Figure \ref{f5}, which indicates that the propagation of shocks supply thermally unstable gas to an extended region.
It seems that oblong grid cells used in lower resolution regions do not show any artificial effect.
In the post-shock region (region where $n_{\rm n}>1.0$ cm$^{-3}$), the average magnetic field strength is $|B|= 11.7$ $\mu$G.

The density of the CNM clumps is typically $n\simeq 3\times10^{1}$ cm$^{-3}$, which corresponds to that of HI clouds.
The shapes of the CNM clumps are filamentary, because the growth of TI due to the motion along the magnetic field is faster than the motion perpendicular to the magnetic field.
We also performed a similar calculation for the ideal MHD equations subject to the same cooling function and thermal conduction with the same initial and boundary conditions.
The result was very similar to the two-fluid case.
In a one-dimensional case, as shown in Inoue, Inutsuka \& Koyama (2007), ambipolar diffusion is effective for the scale of the TI with the presence of microgauss magnetic field.
In a two-dimensional case, however, the mode along the magnetic field dominates the nonlinear growth, and the role of ambipolar diffusion becomes less important at least in the formation regime of the CNM\footnote{The role of ambipolar diffusion may be important through frictional heating in the regime of late time evolution of the two-phase medium as studied in Hennebelle \& Inutsuka (2006)}.
Therefore, even if the converging flows last long, TI cannot pile up material in the direction parallel to the converging flows, and it is difficult to generate a high-density cloud that resembles molecular clouds, since the propagating shocks supply thermally unstable gas to an extended region.

\section{Summary and Discussion}
We have developed an unconditionally stable numerical method for solving the coupling between two fluids, which can adequately solve the evolution of a partially ionized medium from weak-coupling to strong coupling regimes.
By using two-dimensional two-fluid MHD simulations based on this method, we have investigated the dynamical formation process of the CNM in both the magnetized and unmagnetized media.
Our findings are as follows.

In the unmagnetized case, the converging flows of the WNM create a complex slab bound by quasi-standing shocks in which turbulent cloudlets of the CNM are generated as a consequence of TI.
Because the thermal pressure in the slab has to sustain the ram pressure of the flows, the CNM whose pressure is higher than the average pressure of the ISM and whose density is $n\simeq 10^{3}$ cm$^{-3}$, which correspond to the observational characteristics of molecular clouds, is inevitably formed.
Thus, as long as the converging flows continue, the column density of the CNM increases, and a molecular cloud would be generated in the slab.

In the magnetized case, magnetic pressure inside the shocked region can sustain the ram pressure of the flows.
This leads to two important effects on the evolution of the shocked region.
One is that, because the thermal pressure does not have to be high in order to oppose the ram pressure, it decreases due to cooling.
This makes the physical conditions of the resulting CNM similar to those of HI clouds.
Another one is that the shocks continue to propagate, supplying thermally unstable gas to an extended region.
This makes it difficult to generate high-column density clouds in a compact region like molecular clouds.
This is because ambipolar diffusion is not effective globally, and TI cannot pile up material in the direction parallel to the converging flows.

In this paper, as a first step, we examined the situation in which the initial angle between the magnetic field and the velocity of the colliding flows is $90^\circ$, which maximizes the effect of magnetic field.
If the initial angle is small enough, it seems possible to generate the CNM whose physical condition corresponds to molecular clouds, as in the unmagnetized case.
How large is the critical angle that changes the character of the resulting CNM?
In a one-dimensional case, according to Hennebelle \& P\'erault (2000), magnetic pressure prevents condensation, if the initial angle is larger than $20^\circ$-$30^\circ$ with a few microgauss initial field strength (see Fig. 10 of their paper).
In more realistic two-dimensional situation, as we have shown in this paper, the mode of TI along the magnetic field always generates the CNM that corresponds to a HI cloud.
Thus, their conclusion needs to be modified.
However, their result tells that if the initial angle is smaller than their critical angle, magnetic pressure cannot be strong enough to support the ram pressure.
In such a case, the resulting shocked slab would have a physical condition that can generate molecular clouds (precursor of molecular clouds), provided the flows last long.

However, if this conjecture is right, smallness of the critical angle will drastically diminish the opportunity of fast molecular cloud formation directly from the WNM whose time scale is on the order of growth time scale of TI, or cooling time, ($\sim1$-$10$ Myr).
Thus, there is a possibility that most of the molecular clouds are formed from HI clouds, as suggested from observations of clouds in Local Group galaxies (Fukui 2007; Blitz et al. 2007), via more complex processes (e.g., experiences of TI more than once due to successive shocks of supernovae and possibly through the effect of potential well of stellar spiral waves or self-gravity) rather than the direct pile-up of WNM as a consequence of TI.
The diffuculty of the fast formation of molecular clouds in the presence of magnetic field perpendicular to the inflow is also discussed from another point of view in Hartmann et al. (2001).

In order to confirm this possibility, the following studies are necessary:
(1) similar calculations of converging flows varying the angles between converging flows and magnetic field to show the smallness of the critical angle even in multiple dimensions;
and (2) a global study of the magnetized ISM to learn about the typical angle between the mean magnetic field and converging flows or interstellar shocks.
In subsequent papers, we will present studies relevant to these points.

Note that the discussion shown in this section is based on one-/two-dimensional simulations.
However, especially in MHD, dynamical behavior in 3D can be different from that in 2D (see, e.g., Stone \& Gardiner 2007 for Rayleigh-Taylor instability in a magnetized medium).
Thus, we should bear in mind the possibility that the difference due to the dimensional restriction might change the outcome.

\acknowledgments
This work is supported by the Grant-in-Aid for the 21st Century COE ``Center for Diversity and Universality in Physics" from the Ministry of Education, Culture, Sports, Science and Technology (MEXT) of Japan.
Numerical computations were carried out on VPP5000 at the Center for Computational Astrophysics, CfCA, of the National Astronomical Observatory of Japan.
SI is grateful for the hospitality of KITP and interactions with the participants of the program ``Star Formation through Cosmic Time''.
S. I. is supported by Grants-in-Aid (15740118, 16077202, and 18540238) from MEXT of Japan.

\begin{deluxetable}{clc}
\tablewidth{0pt}
\tablecaption{Sources}
\tablehead{
Coefficient &  Note  & Ref.
}
\startdata
$A_{\rm p}$ & $3.4 \times 10^{15}\left( T/8000\mbox{ K} \right)^{0.375}$ cm$^{3}$g$^{-1}$s$^{-1}$ & 1 \\
 & H$^{+}$+H collisions &  \\
$A_{\rm CI\hspace{-.1em}I}$ & $8.5 \times 10^{13}$ cm$^{3}$g$^{-1}$s$^{-1}$ & 2 \\
 & $\rm CI\hspace{-.1em}I$+H collisions &  \\ 
$\kappa$ & $2.5 \times 10^{3} \, T^{0.5}$ erg cm$^{-1}$s$^{-1}$K$^{-1}$ & 3 \\
 & H+H collisions & \\
$\alpha$ & H$^{+}$ and He$^{+}$ recombinations & 4 \\
$\xi$ & $1.0 \times 10^{-16}$ s$^{-1}$ & 5 \\
 & H and He ionizations &  \\
\enddata
\tablerefs{
(1) Glassgold et al. 2005; (2) Draine te al. 1983; (3) Parker 1953; (4) Shapiro \& Kang 1995; (5) Wolfire et al. 1995.
}
\end{deluxetable}
\clearpage

\end{document}